# CAP-RAM: A Charge-Domain In-Memory Computing 6T-SRAM for Accurate and Precision-Programmable CNN Inference

Zhiyu Chen, *Student Member, IEEE*, Zhanghao Yu, *Student Member, IEEE*, Qing Jin,
Yan He, *Student Member, IEEE*, Jingyu Wang, *Member, IEEE*, Sheng Lin, *Student Member, IEEE*,
Dai Li, *Student Member, IEEE*, Yanzhi Wang, *Senior Member, IEEE*,
and Kaiyuan Yang, *Member, IEEE*

*Abstract*— A compact, accurate, and bitwidth-programmable in-memory computing (IMC) static random-access memory (SRAM) macro, named CAP-RAM, is presented for energy-efficient convolutional neural network (CNN) inference. It leverages a novel charge-domain multiply-and-accumulate (MAC) mechanism and circuitry to achieve superior linearity under process variations compared to conventional IMC designs. The adopted semi-parallel architecture efficiently stores filters from multiple CNN layers by sharing eight standard 6T SRAM cells with one charge-domain MAC circuit. Moreover, up to six levels of bit-width of weights with two encoding schemes and eight levels of input activations are supported. A 7-bit charge-injection SAR (ciSAR) analog-to-digital converter (ADC) getting rid of sample and hold (S&H) and input/reference buffers further improves the overall energy efficiency and throughput. A 65-nm prototype validates the excellent linearity and computing accuracy of CAP-RAM. A single 512 × 128 macro stores a complete pruned and quantized CNN model to achieve 98.8% inference accuracy on the MNIST data set and 89.0% on the CIFAR-10 data set, with a 573.4-giga operations per second (GOPS) peak throughput and a 49.4-tera operations per second (TOPS)/W energy efficiency.

## I. INTRODUCTION

DEEP convolutional neural networks (CNNs) achieve unprecedented success in countless artificial intelligence (AI) applications due to their powerful feature extraction capabilities [1]–[3]. In many real-time applications, CNN models are typically pre-trained in the cloud and then deployed in edge devices, such as mobile phones and the Internet-of-Things (IoT) devices, for fast and energy-efficient local inference. Because of the very limited computing resource

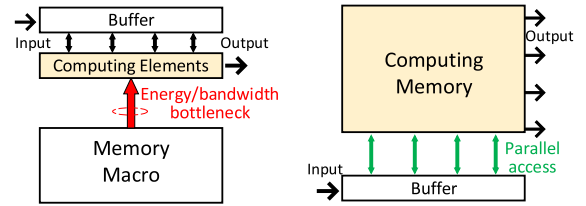

Fig. 1. Conventional and IMC accelerators.

and energy budget, specialized real-time, yet low-power CNN inference hardware is highly desired.

The fundamental and computationally dominant operation of CNNs is convolution. A single convolution step can be expressed by a multiply-and-accumulate (MAC)

$$Y = \sum_{i=1}^{R \times R \times C} W_i X_i \qquad (1)$$

where $Y$, $X$, and $W$ refer to output activations, input activations, and weights, respectively. $R \times R$ represents the kernel size, and $C$ is the number of input channels. It is well known that the energy bottleneck of such computations lies in the overwhelming data movement, rather than arithmetic operations [4]. The energy to access DRAMs and static random-access memories (SRAMs) is approximately $8 \times 10^4$ times and $3 \times 10^3$ higher than that of an 8-bit integer addition in 45 nm [4], leading to the so-called memory wall. The memory walls are particularly severe for data-intensive computing, such as deep learning. State-of-the-art digital CNN accelerators are all optimized for energy-efficient dataflows and reduced memory access, by exploiting data locality and reuse [5]–[7].

To further alleviate the memory walls, emerging non-Von Neumann CNN accelerators that perform computing directly inside the memory by accessing and computing multiple rows in parallel attract significant interests [8]–[19]. In these in-memory computing (IMC) designs, the data movement is significantly reduced, and the read energy is amortized by the parallel access, as shown in Fig. 1. IMC with on-chip SRAMs was first proposed in [20] and first implemented in silicon by Zhang *et al.* [9], which turns on multiple standard 6T SRAM cells at the same time and accumulates current on the bitline to perform energy-efficient MAC computing.

This current-domain IMC technique is then applied to the fully connected layers of binary neural networks [19]. Variants of the current-domain IMC are developed to support multi-bit weights by modulating the pulsewidth of wordline signals [8], and quantized XNOR-nets [21] with customized 12T SRAM cells [13]. A novel 8T SRAM [14] is proposed recently to maintain memory density while preventing read disturbance. While these IMC techniques achieve higher efficiency than their digital counterparts, their application and performance are restricted by the computing inaccuracy caused by process variations. To improve accuracy, charge-domain IMC is developed in [11] and [10], where the analog multiplication is performed on local capacitors and the accumulation is performed by charge sharing among all local capacitors. This charge-domain computation can also be implemented with capacitive coupling to simplify the cell structure [15]. In order to accommodate increasingly diverse CNN structures and bit precisions, reconfigurable in-SRAM computing accelerators [12], [16]–[18], [22] are developed to support multi-bit input activations and weights, pushing IMC toward a universal computing architecture. Recently, advanced techniques, such as sparsity-aware computation [23] and on-chip CNN training [24], are also achieved by IMC.

Since it is difficult to achieve full-precision analog computation inside the memory, precision-configurable IMC architectures work in a bit-serial fashion to support multi-bit computation

$$Y = \sum_{p=1}^{B_W} 2^p \sum_{q=1}^{B_X} 2^q \sum_{i=1}^{R \times R \times C} W_i^p X_i^q \qquad (2)$$

where $B_W$ and $B_X$ are the bitwidth of weights and inputs, respectively. The one-bit MAC operation $\sum_{i=1}^{R \times R \times C} W_i^p X_i^q$ is typically done inside the memory in the analog domain, and the rest of calculation is simply shift-and-add that can be processed by peripheral circuits. To optimize the tradeoff between the computing complexity and throughput, $X_i^q$ can be made multi-bit by adopting digital-to-analog converters (DACs), such as in [11], resulting in a modified (2)

$$Y = \sum_{p=1}^{B_W} 2^p \sum_{q=1}^{\hat{B}_X} 2^{qh} \sum_{i=1}^{R \times R \times C} W_i^p \hat{X}_i^q \qquad (3)$$

where $\hat{X}_i^q$ is an $h$-bit number and $\hat{B}_X = B_X/h$.

While state-of-the-art IMC SRAMs show superior energy efficiency over digital accelerators by leveraging parallelism and fewer memory access and efficient analog computing, they face tradeoffs among computing accuracy, memory density, and precision configurability. With the goal of simultaneously achieving all desired properties of an IMC SRAM, this article presents a *Compact*, *Accurate*, and *Precision-configurable* charge-domain SRAM macro (CAP-RAM) using standard 6T cells. The enabling techniques of the IMC macro include: 1) a compact memory structure supporting lossless charge-domain in-memory computation; 2) a fully reconfigurable semi-parallel computing scheme supporting eight levels of input activations and six levels of weights; and 3) a high-speed and energy-efficient charge-injection SAR (ciSAR) analog-to-digital converter (ADC) avoiding the power-hungry sampling and drivers for reference voltages.

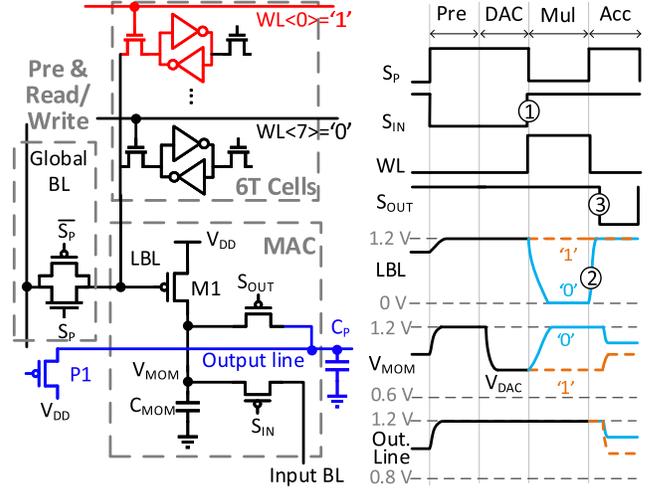

Fig. 2. Proposed 6T charge-domain IMC cluster and operating waveforms.

Moreover, we show that the proposed macro design is compatible with state-of-the-art structured pruning and quantization training methods. The combination of high-density on-chip IMC macro and reduced CNN models promises fully on-chip weight storage and highly efficient inference.

The rest of the article is organized as follows. Section II presents the key ideas and design considerations of the core IMC circuits. Section III covers the details of CAP-RAM implementation, followed by measurement results in Section IV. Section V concludes this article.

## II. PRINCIPLES OF THE PROPOSED CHARGE-DOMAIN COMPUTING WITH 6T SRAMS

### A. Principles of the Core Circuits

The core unit in CAP-RAM is an SRAM cluster for weight storage and charge-domain MAC computing, as shown in Fig. 2. Each cluster consists of: 1) eight standard 6T SRAM cells to store weights; 2) switches and one metal-oxide-metal (MOM) capacitor to perform charge-domain analog MAC computing; and 3) precharge and read/write circuits for normal SRAM operations. For simplicity, the wordline and bitline for the access transistors on the right-hand side of the 6T, which are only used for normal read/write, are omitted in Fig. 2.

Fig. 2 illustrates the operating principles of four IMC phases. In the first reset phase, the local bitline (LBL), local MOM capacitors $C_{MOM}$'s, and the parasitic wire capacitor $C_P$ on the output line are precharged via global BL (GBL), input BL (IBL), and a global PMOS P1, respectively (precharge phase). Next, the 4-bit digital input is converted to a voltage $V_{IN}$ on the IBL by a DAC and sampled on $C_{MOM}$ by turning on $S_{IN}$ (DAC phase). Only one of the eight WLs will be activated so that the stored data in the selected cell control M1 via LBL. The voltage on $C_{MOM}$ is either pulled up to Vdd (equivalent to multiplying by 0) or keeps $V_{IN}$ (multiplying by 1) based on the ON–OFF state of M1, where 1b × 4b analog multiplication is performed (multiplication phase). Notice that the analog operations are all referenced to VDD as logic "0." LBL becomes floating when the accessed cell holds "1." However, the leakage currents from other cells storing "0" cause less than 0.3-mV change on $C_{MOM}$ in the worst case (FF corner, all other cells storing "0"). Finally, addition is performed by turning off M1 and turning on $S_{OUT}$, and hence, the local charge is shared across the $C_P$ and all $C_{MOM}$'s connected to the output line (accumulation phase). The 6T cells in the cluster save 30% area compared with 8T cells and do not suffer from read disturbance because the cells are only connected to LBL with low capacitance (1.5 fF). In our importance sampling-based transient MC failure analysis, the failure rate is out of ten sigmas at the nominal condition. During normal SRAM read/write operations, the switch $S_P$ is turned on to connect all cells directly to GBL, and the rest operations are the same as in standard SRAMs.

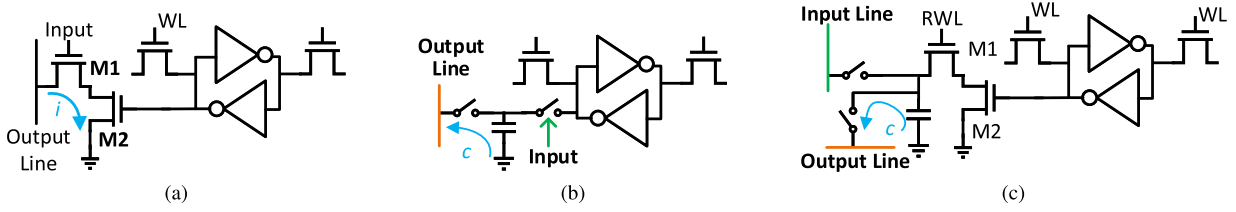

Fig. 3. Existing IMC cell designs: (a) current-domain computation with an 8T SRAM cell, (b) charge-domain computation with wordline input, (c) charge-domain computation with bitline input.

To save energy and area, all switches except $S_P$ are implemented with the only PMOS. This is feasible because the DAC output is already forced to be above half Vdd for linearity considerations. $S_P$ is a transmission gate with split P/N control for SRAM read and write, but only the PMOS is turned on to precharge LBL during computing. One potential problem of the PMOS-only implementation is the charge injection and capacitive coupling effects during switching, which changes the voltage on $C_{MOM}$ due to the small capacitance. Fortunately, the coupling effects caused by ①–③ in Fig. 2 can be mitigated because of their opposite polarities. The coupling error on $C_{MOM}$ is calculated by

$$\Delta V_{CC,\text{MOM}} = \frac{C_{\text{GS}}}{C_{\text{GS}} + C_{\text{MOM}}} V_{DD}. \quad (4)$$

Notice that $S_{OUT}$ not only brings coupling error to the MOM cap but also to the parasitic cap on the output line

$$\Delta V_{CC,P} = \frac{C_{\text{GS}}}{C_{\text{GS}} + C_P/128} V_{DD} \quad (5)$$

where $C_P$ is the total capacitance of the output line. $\Delta V_{CC,\text{MOM}}$ and $\Delta V_{CC,P}$ are constant since the related gate control voltages ($V_{\text{LBL}}$, $S_{\text{OUT}}$, and $S_{\text{IN}}$) are rail-to-rail. On the other hand, charge-injection comes from the residue charge in the channel, so it only happens during switching-off (edge ① and ② in Fig. 2). The error is calculated by

$$\Delta V_{\text{CI}} = \frac{C_{\text{OX}} \text{WL}(V_{DD} - V_{\text{TH}} - V_{\text{MOM}})}{2 C_{\text{MOM}}}. \quad (6)$$

Therefore, the final error that appears on the output line after charge sharing is

$$\Delta V = \Delta V_{\text{ON-OFF}} - \Delta V_{\text{OFF-ON}}$$
$$= \frac{C_P \Delta V_{CC,P} - 128 C_{\text{MOM}} \Delta V_{\text{CI}}}{128 C_{\text{MOM}} + C_P}. \quad (7)$$

As a result, the coupling error on the local MOM cap (4) can be perfectly cancelled, while the charge injection error is partially mitigated by the coupling error on the output line (5). In practice, we tune the size of the three transistors to make sure $\Delta V = 0$ when inputs are all "0000," which cancels the offset of analog computation; 1000 Monte Carlo simulations show that the standard deviation of $\Delta V$ is only 0.6 mV, so the process variation will not significantly affect the cancellation. MOM capacitors of 1.2 fF are implemented for each cluster under tradeoffs between energy and the dynamic range of analog output.

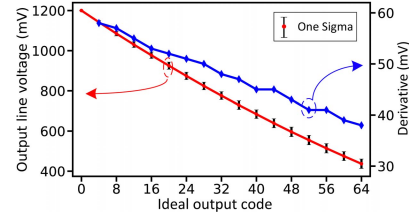

Fig. 4. Simulated linearity of a current-domain 128 × 128 IMC SRAM, with 0.6-V input voltage, 200-ps access time, and 64 activated rows to avoid M1 entering linear region.

### B. Design Analysis and Related Work

*1) Current-Domain Versus Charge-Domain Computing:*
One of the key design choices for IMC macro is the analog computing mechanism. Computing in the current domain was first proposed in [9], which accesses multiple 6T SRAM cells simultaneously and accumulates their discharging current on the bitline. To prevent write disturbance and promote computing accuracy, recent papers introduce 8T cells as storage and computing units [12], [16], as shown in Fig. 3(a).

The main benefit of current-domain computing cells is their simplicity and compatibility with standard SRAM cells. However, it suffers from relatively low computing accuracy due to inherent non-linear $I_{\text{DS}}$ dependence and process variations of the access transistor [M1 in Fig. 3(a)]. The linearity will be even worse if M1 enters the linear region when the bitline voltage becomes low during computing. Fig. 4 shows the simulated linearity and variation (100 Monte Carlo simulations) of a 128 × 128 8T IMC SRAM in the 65-nm CMOS process. The access time of each cell directly affects the linearity, which is set to 200 ps in our simulations. It is limited by the drivers and parasitic capacitance. To meet the saturation condition of M1 ($V_{\text{DS}} > V_{\text{GS}} - V_{\text{TH}}$), the input voltage is set to 600 mV, and 64 rows are activated. To cover the full dynamic range, all cells store "1," and inputs are 1 bit. It is evident from Fig. 4 that the analog computing output is not linear and presents larger variations at lower output voltages.

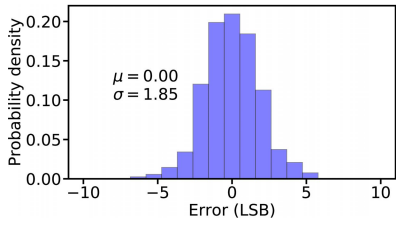

Fig. 5. Simulated histogram of MAC computing error of a current-domain 128 × 128 IMC SRAM.

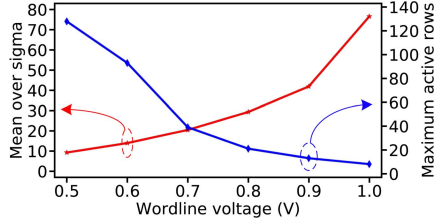

Fig. 6. Simulated variations (mean over sigma value) of the output line voltage in 100 Monte Carlo simulations (when 16 rows are accessed) and maximum active rows, under different input voltages.

The systematic and random non-linearities will both affect computing accuracy. Fig. 5 shows the simulated error distribution with nearly 2-least significant bit (LSB) standard deviation. In this experiment, the same IMC macro above is quantized by an ideal 6-bit ADC model. To cover the full output range better than pure random inputs, the 128 inputs are divided into 16 groups, where the $k$th group has $4k$ "1"s and $64 - 4k$ "0"s at random locations.

The linearity concerns also limit the number of WLs that can be turned on in parallel in current-domain IMC, leading to degraded throughput and efficiency because of fewer parallelism [25]. The simulation results in Fig. 6 depicts the tradeoff between the parallelism and computing accuracy: a higher input voltage reduces the variation of M1 but makes it easier for M1 to enter the linear region, which ultimately restricts the parallelism. The simulations are done on the same design as Fig. 4, with a fast 200-ps access time

In comparison, charge-domain IMC achieves better computing accuracy and higher parallelism [10], [11]. The computation [as shown in Fig. 3(b) and (c)] is performed on capacitors, which has much less variation than the current of minimum-sized access transistors. Meanwhile, the charge-sharing based operation is not affected by transistors' operating regions, and therefore, a greater number of cells can be turned on together for higher throughput and efficiency gain. It is clear that no significant linearity degradation is observed even in measurements of CAP-RAM (see Section IV-A1). Therefore, charge-domain computing is a clear choice for accurate and higher-precision IMCs.

*2) Wordline Versus Bitline Inputs:* Fig. 3 depicts three categories of designs with two approaches to supply the inputs for convolutions, wordline, and bitline inputs [10]–[12]. Current-domain IMC is typically done with wordline inputs. The pulsewidth modulated WL signal can represent multi-bit inputs.

Charge-domain cells support both approaches. In cells with wordline inputs [see Fig. 3(b)], a logic AND is performed between the input and stored value with the output switch off. Then, the switch is turned on so that the charge is shared across all the capacitors connected to the output line. This is a simple and efficient scheme for binary inputs but requires high precision circuits to support multi-bit inputs using pulsewidth modulation or current modulation because the sampling capacitors are tiny (a few fFs) for energy considerations. On the other hand, the IMC with bitline input refers to sampling the input signals on a local capacitor [see Fig. 3(c)]. This scheme will support multi-bit inputs. When the input is an $h$-bit digital signal, it will first be converted to voltage $V_{IN}$ on the input line by a DAC. $h$-bit × 1-bit multiplication is performed by closing RWL switch, similar to Fig. 3(b), and the output line switches are later turned on to finish accumulation. As indicated in (3), a $h$-bit input architecture achieves nearly $h$ times throughput improvement over the pure bit-serial scheme that needs $h$ loops to perform the same operations. Thus, CAP-RAM (Fig. 2) adopts bitline inputs to support higher throughput.

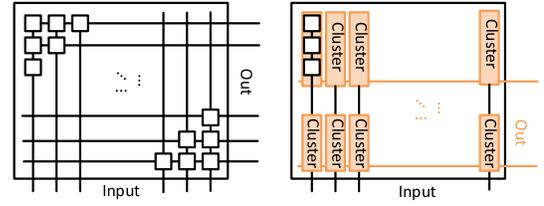

Fig. 7. Comparison between the fully parallel structure and the clustering structure.

*3) Clustering Structure:* Conventionally, IMC SRAMs are expected to activate all rows simultaneously to maximize energy efficiency and compute density, while CAP-RAM groups several 6T cells and one analog computing module into a cluster (as shown in Fig. 7), and only one of those cells will be selected at each operation. This is the result of a design compromise to amortize the large bitcells needed for the charge domain in-SRAM computing. Larger bit cells not only reduce compute density but also increase energy and delay over an ideal fully parallel IMC 6T-SRAM. On the other hand, as discussed in Section I, analog MAC with multi-bit inputs could linearly increase the throughput and energy efficiency over bit-by-bit serial computing [15], [26], which is leveraged in CAP-RAM together with the clustering structure to offer higher macro-level compute density than fully parallel macros. For instance, state-of-the-art charge-based serial computing cells, even for bit-by-bit serial computing, are about two to three times the area of a logic-rule 6T SRAM cell [15], [16], [26].

Comparatively, the standard 6T SRAM cell is used in CAP-RAM. Our implementation is in logic rule, but "push-rule" cells with ∼50% less cell area can be easily adopted. The switching circuit is around three times the size of a logic-rule 6T cell. The area overhead of the switching circuit can be greatly amortized by the clustering structure. For example, a CAP-RAM cluster of three non-push-rule cells will take the same area as two or three cells doing bit-by-bit serial computation. If a 4-bit input × 1-bit weight MAC is performed within the same sized array, CAP-RAM will provide the highest total # of operations (bitwise multiply and add) per cycle. In this iso-area comparison, the CAP-RAM will simultaneously achieve the highest compute density, throughput, and also higher weight storage density. If push-rule layouts are considered, the gaps will become even larger because custom cells are unlikely to take full advantage of push rules. Previous multi-bit MAC computing IMC SRAM [11] also adopts the clustering structure, but the use of 10T SRAM cells and more transistors in the switching circuits lead to ∼50% more cell area and less compute density than CAP-RAM.

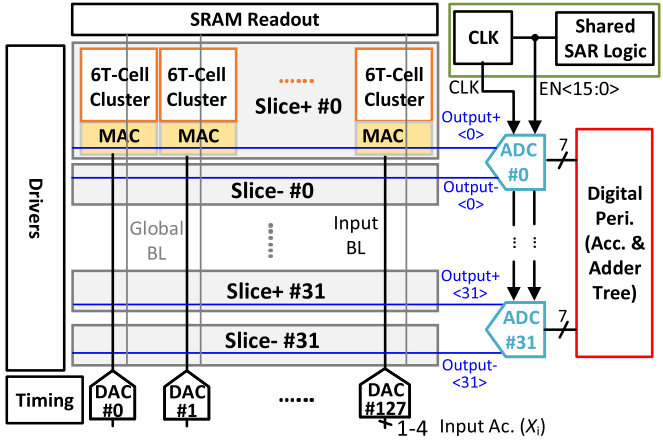

Fig. 8. System diagram of the proposed charge-domain IMC architecture.

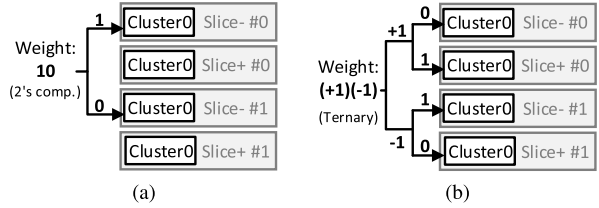

Fig. 9. Illustration of the mapping of (a) 2's complement encoding and (b) ternary encoding.

Meanwhile, the clustering IMC macro enables a candidate architecture for edge inference, where IMC macros (i.e., PE) stores the entire model. Compared to the classical architecture with large on-chip buffer or embedded non-volatile memory to store weights, CAP-RAM potentially offers comparable storage density as an SRAM buffer and eliminates latency and energy associated with weight loading. When CAP-RAMs are used for such architecture, each cluster can be used to store different CNN layers not executed in parallel. This will further alleviate the penalty on parallelism due to clustering. It is worth mentioning that fully parallelism of all MAC operations is possible with inter-layer pipelining [27], [28], and the imbalanced speed/throughput of each pipeline stage can be solved by mapping techniques, such as replication. However, such architectures also come with high overall power and pipelining overhead and also requires the pipelines to be filled by streaming inputs, both of which may not be acceptable in many edge applications. To this end, the proposed clustering structure offers a tuning knob for the tradeoffs between the weight storage density and the compute density. Generally, the greater the number of cells in a cluster, the lower the compute density, but the higher the storage density. For specific applications, the CAP-RAM based hardware can be moved on the optimization plane through co-design of cluster size and mapping strategy. In our prototype, eight-cell clusters are employed to maximize on-chip storage density and allow efficient layer-by-layer execution of LeNet and ResNet models.

## III. IMPLEMENTATION

The proposed CAP-RAM macro consists of (see Fig. 8): a 6T SRAM array with charge-domain IMC switches, current-steering DACs for each column, ciSAR ADCs for each pair of the slice, and digital peripherals after ADCs. In the prototype, a 512 × 128 memory divided into 32 pairs of slices is implemented. Each slice consists of 128 clusters and performs one MAC computation on the output line, which is connected to one of the differential inputs of an ADC.

IMC and analog-to-digital conversion are done serially in one global clock cycle (70 MHz). During IMC, 128 4-bit inputs are first converted to voltages via 128 DACs and transmitted to 64 clusters on the input line. The output line voltage is then quantized to digital by a 7-bit SAR ADC. The digital codes are registered and sent to the digital periphery, which consists of serial accumulators and adder trees to accumulate partial sums and combines to support programmable input and weight precision.

### A. Reconfigurable Support of Two Data Encoding Schemes

The proposed architecture supports both 2's complement and ternary encodings of weights, which are most common in CNN quantizations. This is achieved by grouping two neighboring slices into a pair ("+" and "−" slices in Fig. 8) and connecting their outputs to differential ADC inputs. The computation and quantization can be programed to single-ended or differential modes to support the two encoding schemes in different CNN layers.

The 2's complement encoded data are binary and, thus, only require a single SRAM cell for each bit. To conduct IMC for $k$-bit 2's complement weights, they will be stored in the same column but $k$ neighboring "+" or "−" slices, as shown in Fig. 9(a). The two slices in each pair alternately turn on with the ADC set to single-input mode. To obtain the final MAC result, $k$ consecutive ADC outputs will be shifted and summed up, which are done in the digital periphery. Since the most significant bit (MSB) is negative, the sign of the corresponding partial sum needs to be reversed. The 2's complement encoding is preferred for storage efficiency because $k$ bit cells contain $k$-bit information.

On the other hand, each ternary encoding unit $(-1/0/+1)$ requires two binary SRAM cells. For example, $6 = 2^3 \times (+1) + 2^2 \times (-1) + 2^1 \times (+1) + 2^0 \times (0)$. As shown in Fig. 9(b), two cells with the same index within a slice pair store one encoding unit: "01" for "+1," "10" for "−1," and "00" for "0." The ADCs run in differential-input mode and, hence, naturally perform the subtraction of the positive and negative results without extra processing. In general, to store $k$-bit information, $2k-2$ bitcells are needed for ternary encoding, but only $k-1$ ADCs are required. This is particularly useful for ternary neural networks with only $-1/0/+1$ weights, because they require the same memory footprint as 2-bit 2's complement weights, but saves one ADC for each MAC.

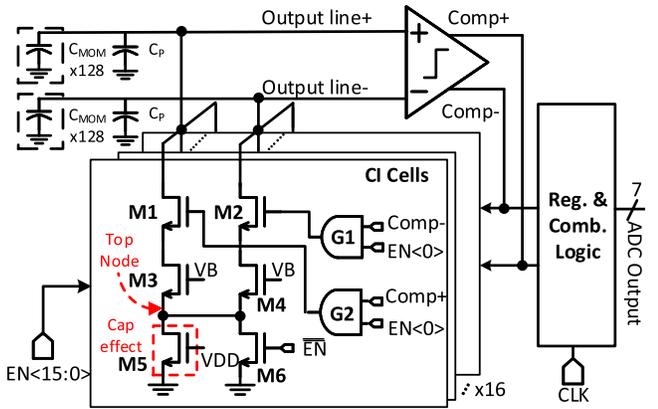

Fig. 10. Diagram of the ciSAR ADC. The global SAR control logic and clock are shared by 32 ADCs.

### B. Compact and Driver-Less ciSAR ADC

ADCs are critical in the design of IMC macros because of the stringent area and energy constraints. Flash ADCs are widely used for low-precision IMC [13], [15], [29] due to their simplicity, but its area scales quickly with a resolution. A serial ADC [11], which uses a single dummy cell as a reference, achieves 7-bit resolution and high area-efficiency, but the serial nature results in low throughput.

SAR ADCs offer balanced area, throughput, and energy in the 5–8-bit precision range that is most common for CNN quantizations. For example, a 5-bit SAR ADC is adopted in [12]. However, conventional capacitor-based SAR ADCs face several challenges in IMC applications. First, the capacitive DAC is large so that the total area of ADCs can be as large as that of the SRAM array [12]. Second, powerful reference voltage buffers with large output currents are necessary to drive the bottom plates of all capacitive DACs in the same macro. Finally, in charge-domain IMC, the analog computing output is in charge and, thus, requires an input analog buffer to drive sample and hold (S&H) circuits. It is worth mentioning that direct charge sharing from local $C_{MOM}$'s to the top plate of the capacitor array (about 160 fF for a 7-bit ADC) is prohibitive due to significant dynamic range loss. Assuming $C_{MOM} = 1.2$ fF and $C_P = 80$ fF, the input range reduces to $1.2 \times 128/(1.2 \times 128 + 80 + 160) = 39\%$ for a 7-bit ADC and further drops to 28% for an 8-bit ADC.

This work exploits ciSAR ADCs, first proposed in [30], to address the challenges above and meet the specific requirements of CAP-RAM. Capacitors in the DACs are replaced by transistors with a long channel length (see M5 in Fig. 10) in the charge-injection cells (CI cells), which takes much less area than unit capacitors in conventional SAR ADCs. M5 behaves like a capacitor because it can store charge in the channel. Despite the compact area (429 $\mu m^2$), ciSAR ADC achieves 6.85 effective number of bits (ENOB) in transient noise simulations. Unlike conventional SAR ADCs that are referenced by voltages connected to the bottom plate of the capacitive DAC, ciSAR ADC controls the conversion step through a bias voltage VB controlling the current of M3, and therefore, no powerful reference driver is required. Moreover, in conventional SAR ADCs, the input voltage is sampled on the DAC's capacitor with an input buffer and S&H circuits. Because the MAC output is sampled on a capacitor, an input buffer with a sufficient slew rate is necessary, which consumes significant energy and area. Comparatively, the ciSAR ADC only requires the signal to be sampled on a separate large capacitor. Thus, no extra S&H circuits are necessary for CAP-RAM because the analog MAC outputs are already computed and stored on 128 local $C_{MOM}$'s and $C_P$ after charge sharing.

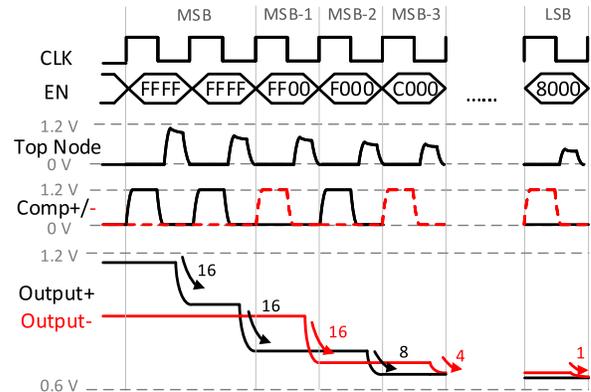

Fig. 11. Operating waveforms of the ciSAR ADC.

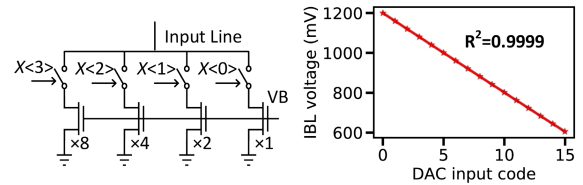

Fig. 12. Diagram of the current-steering DAC and the simulated linearity.

Since CI cells can only deduct the charge from the main capacitor, the SAR logic follows the monotonic switching procedure proposed by Liu *et al.* [31]. Part of the SAR control module is shared among 32 ADCs so that the area is amortized. The MSB of the 7-bit differential ADC is processed by discharging all 16 CI cells twice in order to save the number of CI cells (see Fig. 11). At each SAR step, either M1 or M2 will be turned on based on the comparison result. The charge is then shared from $C_{MOM}$'s and $C_P$ to the top node of the CI cell. This charge transferring is fast because M5 has a faster settling speed than MOM capacitors [30]. M6 is used to reset the "Top Node" before each conversion cycle. CI cells for different SAR step are unary coded to mitigate mismatch. The ADC can be easily switched to the single-input mode by setting $S_{OUT}$ of the disabled slice to "0" and precharging its $C_{MOM}$'s and $C_P$ to Vdd. The rest of the operations are identical to the differential case. Naturally, the single-ended ADC has a 6-bit resolution because the sign bit of the 7-bit output code will be discarded.

### C. Current-Steering DAC

A basic current DAC using biased transistors is used to generate the 4-bit input signal for each column, as shown in Fig. 12. The input bitlines are first precharged and then discharged by the DAC. The biased transistors are up-sized to reduce the process variation and the effect of channel length modulation yet match the pitch of a 6T cell. The 4-bit input digital code controls the binary sized current paths.

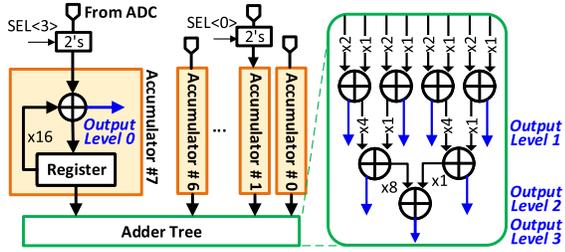

Fig. 13. Diagram of the adder-tree-based digital processing periphery.

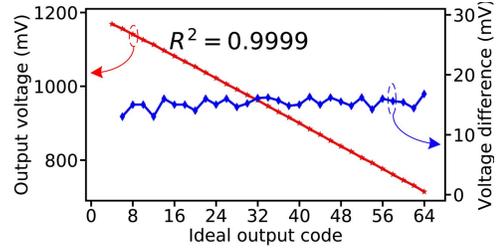

Fig. 15. Measured computing linearity of charge-domain IMC.

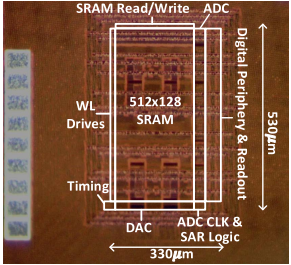

Fig. 14. Micrograph of the prototype chip.

Monte Carlo simulations show that the standard deviation of the input line voltage is only around 1.56 mV in the worst scenario. In order for the biased transistor to remain in the saturation region, its biasing voltage and discharging time are set to keep the bitline voltage above 600 mV. Simulation results demonstrate excellent linearity with $R^2 = 0.9999$ (see Fig. 12).

### D. Digital Processing Periphery

The digital periphery after ADCs is responsible for shifting and adding the partial sums based on (3). As shown in Fig. 13, the ADC output code is first passed through a 2's complement transformation module where the sign of MSB's partial sum will be reversed when 2's complement encoding is used. Notice that the "+1" operation in the transformation is done by the adder inside the accumulator. To support up to 8-bit input activations, the system takes two global cycles to add the two partial sums in the accumulator, while inputs of less than 4 bits only require one clock cycle. The adder tree sums up the results of each accumulator to combine the partial MAC results calculated with different bits of the weights. The adder tree is programmable with four levels of outputs (see Fig. 13) to support six weight bitwidths (1/2/3/4/5/8). At output level $i$, $2^i$ slices are combined, so $2^i$-bit weights are supported by 2's complement encoding (1/2/4/8) and $(2^{i-1} + 1)$-bit weights ($i \neq 0$) by ternary encoding (2/3/5).

Compared to the switch capacitor-based analog shift-adders like [8], the adder tree-based approach has the reconfigurability to process multiple bitwidth of weights. More importantly, analog shift-adders require the voltage to be sampled on separate switch capacitors by charge sharing, but this operation will significantly reduce the input dynamic range of ADCs since the analog computing voltage has already been sampled on $C_{MOM}$s and $C_P$s.

## IV. MEASUREMENT RESULTS

A prototype chip, shown in Fig. 14, is fabricated in a 65-nm LP CMOS process. The total area of CAP-RAM macro is 0.179 mm$^2$. The 8KB SRAM array occupies 62.6% of the area, and the amortized bitcell area is about 38% larger than that of the standard 6T cell. The ciSAR ADCs and current DACs only occupy 8.2% and 3.6% of the total macro area, respectively; 5.8% of the area is used by control signal drivers and read/write circuitry, and 13.7% is occupied by the digital processing periphery and readout registers. In all experiments, the test chip is interfaced with a host PC through digital I/O devices.

### A. Linearity

The accuracy of analog IMC is largely decided by the linearity of each component in the computing pipeline, i.e., DAC, multiplication and addition, and ADC. We thoroughly evaluate the linearity of each component and propose a post-silicon calibration approach to enhance the linearity.

*1) Linearity of Charge-Domain MAC Computing:* Fig. 15 shows the measured output line voltage of one slice under different input patterns. The ideal output code refers to the exact MAC computing results quantized by an ideal 6-bit ADC. To isolate the impacts of the non-ideality of DACs, the input code of each DAC is either "0000" or "1111." Due to the charge-sharing mechanism, almost perfect linearity is achieved with a 0.9999 $R^2$ value.

*2) Linearity of ADCs:* The linearity of the ADC is measured indirectly by controlling the MAC inputs. Given the high linearity of charge-domain IMC, a reasonable estimation of ADC's linearity can be obtained. The ADCs operate in 7-bit differential mode, but only half of the ADC dynamic range (6 bit) is used because the computing results with a certain weight pattern only span half of the full range. As shown in Fig. 16(a), the 32 slices in one chip show offset and gain variations. Comparators' offset in ADCs is the main source of offset. The gain variations are caused by variations of sampled charge, charge sharing ratio, and ADC's conversion step among different slices. The sampled charge $\left(= \sum_{i=1}^{128} \times C_{MOM} V_{IN}{}^i\right)$ of different slices varies due to mismatch of $C_{MOM}$'s, while mismatch of $C_{MOM}$ to $C_P$ ratio results in different charge sharing ratio. Inconsistent ADC's conversion step is caused by the mismatch of CI cells and input capacitors ($128 \times C_{MOM} + C_P$). The mismatch observed is expected due to the area and power constraints in the implementation. Many known analog techniques can mitigate the variations, but, in this work, we focus on using digital calibration.

*3) Calibration:* We propose a two-step low-cost calibration to mitigate the variations and improve linearity. The calibration utilizes the measured data from Fig. 16(a). First, the 32 curves are linear fitted with $y_i = k_i x + b_i$, where $y_i$ is the measured output of the $i$th ADC and $x$ is the ideal output. Each raw output code $\hat{y}_i$ is calibrated by $(\hat{y}_i - b_i)/k_i$ to remove the offset and gain error. Furthermore, a master curve, which is widely used in low-power analog applications, such as temperature sensors, can be applied to alleviate the systematic nonlinearity. The final two-step calibrated MAC result with better linearity is shown in Fig. 16(b). Integrating the calibration module on the chip requires some extra arithmetic logic in the periphery, but the area and energy will be much smaller than that of existing convolution and batch normalization steps.

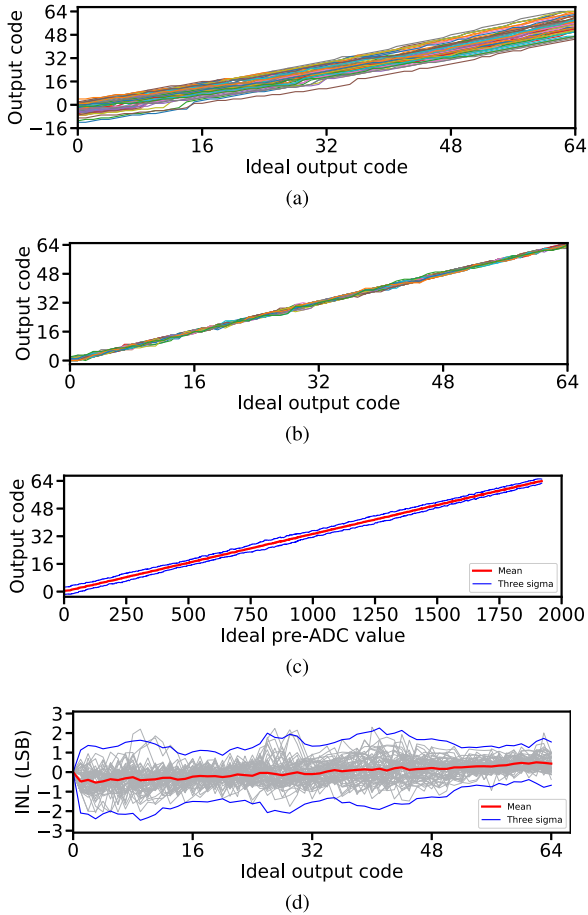

Fig. 16. Measured linearity of the CAP-RAM arrays in two prototype chips. (a) Raw transfer curves of 64 ADCs. (b) Transfer curves of 64 ADCs after two-step calibration. (c) Linearity of the complete analog computing pipeline over 64 slices. (d) INLs of 64 slices.

*4) Linearity of the Analog Computing Slices:* We further analyze the linearity of the complete analog computing chain by including DAC's nonideality. Instead of examining the linearity of a single analog chain as in most ADC studies, it is more important to examine the distribution of INLs and transfer curves across different computing chains and different chips for IMC applications. Therefore, the mean and three-sigma spread of the transfer curves and INLs of 64 slices in two prototype chips are plotted in Fig. 16(c) and (d). Fig. 16(c) is obtained by sweeping the pre-ADC inputs from 0 to 1920. Fig. 16(d) indicates that the largest linearity error of the system is expected to be less than two LSBs. Note that the nonlinearity here includes contributions from the DAC, the analog computation, and the ADC. The measurement also confirms that DAC's nonideality does not significantly degrade the system's linearity, validating the analysis in Section III-C.

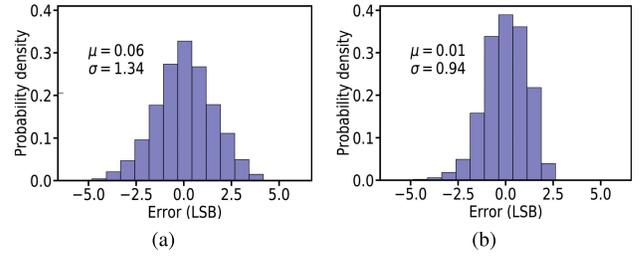

Fig. 17. Measured system error distribution over 524 288 samples (a) after linear-fitting and (b) after linear-fitting and master curve calibration with noise filtered out.

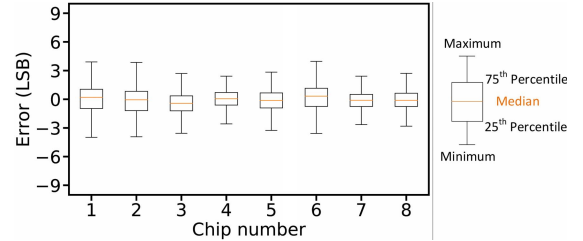

Fig. 18. Measured error distribution of eight prototype chips.

### B. Computing Accuracy

Nonidealities described above (analog computing error, DAC nonlinearity, ADC nonlinearity, and offset/gain variations), together with thermal noise, decide the final computing errors. In addition to linearity tests, we directly assess MAC computing errors by feeding random sets of inputs to the system and comparing the outputs against expected ones.

*1) Error Distribution:* If the inputs are uniformly sampled, the MAC outputs will mostly appear around the center in the dynamic range as a result of the central limit theorem. To alleviate such bias in the measured error distribution, 16 random input sets with different distributions are used. In the $k$th set, 64 different input patterns are randomly sampled from $N(k-1, 2)$. In total, 524 288 ($16 \times 64 \times 32 \times 16$) samples are collected for Fig. 17 because there are 32 ADCs and every measurement is repeated 16 times. Fig. 17(a) shows the error distribution after the system is calibrated by the linear fitting. The spread is further reduced when noise is filtered by averaging the outputs of multiple runs with the same inputs, and the master curve calibration is performed, as shown in Fig. 17(b). Compared with the simulated error distribution of the current-domain system in Section II-B, the error of CAP-RAM is still smaller despite the ideal ADC and DAC (1-bit) assumption. Fig. 18 shows the error distribution over eight chips.

*2) Random Errors:* The thermal noise in ADCs and DACs is another source of computing errors. Fig. 19(a) shows the rms errors of one ADC over the entire input range. The average rms is 0.35 LSB. Spikes can be observed when the input voltage of the ADC is close to the transition threshold. The variation of rms noise across 32 ADCs in the same macro is shown in Fig. 19(b). This noise level is acceptable for ADCs targeting low power and compact area, but it can be further improved in case of relaxed area and power requirements.

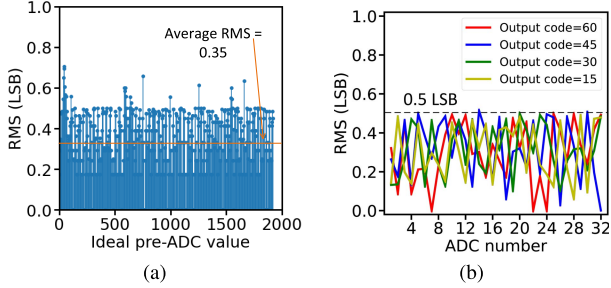

Fig. 19. Measured rms error (a) of one ADC over pre-ADC values (i.e., analog MAC outputs) and (b) 32 ADCs at four pre-ADC values; rms errors are tested over 128 runs with repeated inputs.

*3) Inference on MNIST:* The system achieves 98.8% inference accuracy for the MNIST data set using compressed LeNet-5 [32], which is identical to the software baseline with 500 images tested. The CNN is specifically trained for CAP-RAM and pruned by 95.6% over the baseline model using the alternating direction method of multipliers - neural network (ADMM-NN) framework [33]. The pruned model has only 19 149 parameters. Therefore, the whole network can be stored in a single CAP-RAM macro. The mapping strategy of 2's complement encoded weights follows three rules.

1) The bits of weight are assigned to the same column but neighboring "+" or "−" slices, as described in Section III-A.
2) Within a single filter, the bits from different weights but the same bit location are mapped to the same row. Different filters are mapped to different slices. If the filter size (R × R × C) is larger than the row size, one filter will be mapped as multiple filters.
3) One layer occupies one row in each slice. If the rows out of 64 slices cannot accommodate a layer, a new row in each slice will be occupied in the same way.

The mapping strategy for ternary encoding is similar, except that its encoding unit (+1/0/−1) utilizes two bitcells, and they are paired in the same column and adjacent slices. Table I summarizes the different hardware utilization (rows per slice) for the four layers. In particular, F5 occupies four rows in each slice but only takes four cycles to process in this mapping. This is trivial compared to convolutional layers (576 cycles for C1), and a huge amount of area is saved due to the clustering structure. The system is in the single-ended mode for C1 for efficient storage, while the other layers use differential mode because they have ternary weights. It is worth mentioning that only linear fitting calibration is applied here, but one can also apply both calibration steps in case the CNN model is more sensitive to computing errors.

*4) Inference on CIFAR-10:* A quantized ResNet-20 [3] (see Table II) is deployed on CAP-RAM for the CIFAR-10 data set. The same mapping strategy will be applied to multiple macros when one macro cannot hold a whole layer (layers 15–19 occupy two macros). One of the challenges of the model is to mitigate the effect of quantization errors because the output levels of ADCs (6 bit in single mode) are much smaller than the voltage levels (about 11 bit) of inputs. Therefore, a quantization-aware training approach is utilized to tolerate the quantization errors. CAP-RAM achieves 89.0% inference accuracy with 500 images tested and shows 1.6% degradation compared to the software baseline. Compared with recent IMC architectures, Jia et al. [26] show higher accuracy since it uses 8-bit ADCs and utilizes VGG-like network with 7.04 times more operations than ResNet-20. In [12] and [22], 90.42% and 92.02% accuracies are achieved with ResNet-20, but parallelism is sacrificed to avoid quantization errors. In both designs, a 5-bit ADC only processes less than 16 rows of analog computing, limiting the throughput and efficiency gains.

TABLE I
PRUNED AND QUANTIZED LENET-5 STRUCTURE AND MAPPING

|  | C1 | C3 | FC5 | FC6 |
|---|---|---|---|---|
| **Kernel Size** | 5 × 5 | 5 × 5 | \ | \ |
| **# Kernels** | 5 | 16 | 64 | 10 |
| **Input bitwidth** | 8 | 4 | 4 | 4 |
| **Weight bitwidth** | 4 | 2 (ternary) | 2 (ternary) | 2 (ternary) |
| **ADC Mode** | Single | Differential | Differential | Differential |
| **Occupied Rows**[a] | 1 | 1 | 4 | 1 |

[a] the number of occupied rows in each slice

TABLE II
QUANTIZED RESNET-20 STRUCTURE AND MAPPING

|  | Layer 1 | Layer 2-7 | Layer 8-13 | Layer 14-19 |
|---|---|---|---|---|
| **Kernel Size** | 3 × 3 | | | |
| **# Kernels** | 28 | 28 | 28 | 56 |
| **Input bitwidth** | 8 | 4 | 4 | 4 |
| **Weight bitwidth** | 4 | | | |
| **ADC Mode** | Single | | | |
| **Occupied Rows** | 2 | 4 | 4 | 8 |

*C. Energy and Throughput*

The CAP-RAM prototype operates at 70 MHz and achieves 573.4-giga operations per second (GOPS) peak throughput and 3.4 tera operations per second (TOPS)/mm$^2$ for convolution with 4-bit inputs and binary (1b) or ternary (2b) weights at 1.2-V supply voltage and 25 °C. The compute density is higher than state-of-the-art programmable charge-domain IMC [26], [34] despite CAP-RAM has 1-to-8 row parallelism and multi-bit analog computing circuits. This is because: 1) the ciSAR ADC has high speed and 2) the clustered 6T cells have a highly compact layout, so the reduced parallelism will not degrade the compute density significantly. Although C3SRAM [15] has higher compute density (20.2 TOPS/mm$^2$), yet one operation in C3SRAM is only 1'b by 1'b addition/multiplication and a flash ADC with only 11 levels is used for quantization. Since the throughput of bit-serial architectures naturally scales with the bitwidth, a bitwise compute density (1 OP = 1'b by 1'b addition or multiplication) is defined to make apple-to-apple comparisons, and CAP-RAM becomes the best (27.2 TOPS/mm$^2$) in this metric.

The whole system consumes 11.62 mW with random 4-bit inputs and 1-bit weights. The SRAM array, DACs, and timing controller consume 3.60 mW in the single-ended mode and 6.35 mW in the differential mode with all 32 slices turned on. The power consumption of the 32 ADCs and the shared control and timing module is 7.56 mW. For the digital periphery, the accumulators and 2's complement modules take 0.78mW at the accumulation mode and 0.46 mW at the single-cycle mode, while the adder tree consumes 0.04/0.10/0.19 mW at the output level 1/2/3. Different from the fully bit-serial architectures, CAP-RAM's energy efficiency is based on MAC computation with 4'b inputs. To achieve this, more rowwise control signals and 128 4'b DACs are involved. Similar to the definition above, CAP-RAM becomes more competitive in the bitwise energy efficiency. More importantly, there exists a tradeoff between storage density and energy efficiency. The target of CAP-RAM is not to achieve the highest energy efficiency but to design a compact, accurate, and programmable architecture while maintaining competitive energy efficiency. The detailed performance comparison is summarized in Table III.

TABLE III
PERFORMANCE SUMMARY OF CAP-RAM AND COMPARISON WITH STATE-OF-THE-ART IN-MEMORY COMPUTING SRAMS

| | This Work | JSSC'19 [12] | SSCL'19 [15] | JSSC'18 [11] | ISSCC'18 [35] | JSSC'20 [26] | JSSC'19 [34] | ISSCC'20 [22] |
|---|---|---|---|---|---|---|---|---|
| **Technology (nm)** | 65 | 55 | 65 | 65 | 65 | 65 | 65 | 28 |
| **Supported Input Bitwidth** | 1-8 | 1-4 | 1 | 1-7 | 8 | 8 | 1 | 1-8 |
| **IMC Input Bitwidth** | 4 | 2 | 1 | 7 | 4 | 1 | 1 | 2 |
| **Encoding Scheme** | Ternary/2's | 2's | Binary | Binary | 1's | 2's | Binary | 2's |
| **Weight Bitwidth** | 1/2/3/4/5/8 | 2/5 | 1 | 2 | 8 | 1/2/4/8 | 1 | 4/8 |
| **ADC Resolution** | 7 | 5 | 4 | 7 | 8 | 8 | 1 | 5 |
| **Cell Structure** | 6T | Twin 8T | 8T | 10T | 6T | 10T | 10T | 6T |
| **Computing Mechanism** | Charge | Current | Charge | Charge | Current | Charge | Charge | Current |
| **Array Size** | 512 × 128 | 64 × 60 | 256 × 64 | 256 × 64 | 512 × 256 | 2304 × 256 | 2.4 Mb | 512 × 128 |
| **Area/Cell ($\mu m^2$)** | 2.6 | 12.2 | 4.9 | 3.8 | 6.18 | 14.5 | 5.25 | N/A |
| **Effective Area/Cell[a] ($\mu m^2$)** | 20.8 | 43.4 | 4.9 | 61.5 | 791 | 14.5 | 5.25 | N/A |
| **Peak Throughput (GOPS[c])** | 573.4 (I:4b W:2b) | 67.5 (I:1b W:2b) | 1638 (I:1b W:1b) | 8 (I:7b W:1b) | 16 (I:8b W:8b) | 2156 (I:1b W:1b) | 18876 (I:1b W:1b) | 62.4 (I:4b W:4b) |
| **Compute Density (TOPS/$mm^2$)** | 3.4 (I:4b W:2b) | 1.8 (I:1b W:2b) | 20.2 (I:1b W:1b) | 0.13 (I:7b W:1b) | 0.02 (I:8b W:8b) | 0.6 (I:1b W:1b) | 1.5 (I:1b W:1b) | N/A |
| **Energy Efficiency (TOPS/W)** | 49.4 (I:4b W:1b) | 72.11 (I:1b W:2b) | 671.5 (I:1b W:1b) | 40.3 (I:7b W:1b) | 6.25 (I:8b W:8b) | 192 (I:1b W:1b) | 866 (I:1b W:1b) | 58.1 (I:4b W:4b) |
| **Bitwise Throughput[c]** | 4587.2 (I:4b W:2b) | 375 (I:4b W:5b) | 1638 (I:1b W:1b) | 56 (I:7b W:1b) | 1024 (I:8b W:8b) | 2156 (I:1b W:1b) | 18876 (I:1b W:1b) | 998.4 (I:4b W:4b) |
| **Bitwise Compute Density[c]** | 27.2 (I:4b W:2b) | 3.6 (I:1b W:2b) | 20.2 (I:1b W:1b) | 0.91 (I:7b W:1b) | 1.28 (I:8b W:8b) | 0.6 (I:1b W:1b) | 1.5 (I:1b W:1b) | N/A |
| **Bitwise Energy Efficiency[c]** | 319.2 (I:4b W:2b) | 367.4 (I:4b W:5b) | 671.5 (I:1b W:1b) | 282.1 (I:7b W:1b) | 400 (I:8b W:8b) | 192 (I:1b W:1b) | 866 (I:1b W:1b) | 929.6 (I:4b W:4b) |
| **MNIST Accuracy** | 98.8% | 99.52% | 98.31% | 98.3% | N/A | N/A | 98.6% | N/A |
| **CIFAR-10 Accuracy** | 89.0% | 90.42% | 85.5% | N/A | N/A | 92.4% | 83.27% | 92.02% |

[a]Effective Area/ Cell = $\frac{\text{Total Area}}{\text{\# Active Cells in IMC}}$   [b]1OP=1 addition or 1 multiplication
[c]Bitwise Metric = Metric×IMC Input Bitwidth×Weight Bitwidth

## V. CONCLUSION

In summary, this work presents and demonstrates a charge-domain IMC SRAM macro with 6T cells. The charge-sharing mechanism ensures good accuracy, while the semi-parallel architecture provides best-in-class weight storage density. Meanwhile, the digital processing periphery provides input/weight bitwidth configurability, and a ciSAR ADC specifically designed for CAP-RAM further boosts the energy and area performance. A 65-nm prototype demonstrates excellent computing linearity and accuracy. The pruned and quantized LeNet-5 and ResNet-20 are mapped to CAP-RAM macros, which achieve 98.8% inference accuracy on MNIST and 89.0% on CIFAR-10, respectively. The system achieves 49.3 TOPS/W energy efficiency and 573.4-GOPS throughput.

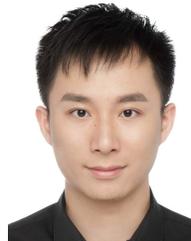

**Zhiyu Chen** (Student Member, IEEE) received the B.E. degree in electrical engineering from Nanjing University, Nanjing, China, in 2018. He is currently pursuing the Ph.D. degree in electrical and computer engineering at Rice University, Houston, TX, USA.

His research interests include digital and mixed-signal circuit design for machine learning accelerators.

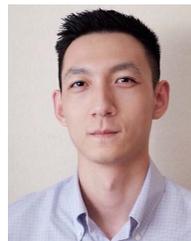

**Zhanghao Yu** (Student Member, IEEE) received the B.E. degree in integrated circuit design and integrated system from the University of Electronic Science and Technology of China, Chengdu, China, in 2016 and the M.S. degree in electrical engineering from the University of Southern California, Los Angeles, CA, USA, in 2018. He is currently pursuing the Ph.D. degree in electrical and computer engineering at Rice University, Houston, TX, USA.

His current research interests include analog and mixed-signal integrated circuits design for power management, bio-electronics, and security.

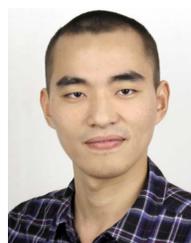

**Qing Jin** received the M.S. degree in computer engineering from Texas A&M University, College Station, TX, USA, in 2018 and the B.S. and M.S. degrees in microelectronics from Nankai University, Tianjin, China, in 2009 and 2012, respectively.

He was working as a Research Assistant with Tsinghua University, Beijing, China, between 2010 and 2012. From 2013 to 2017, he was working with the School of Microelectronics, Xi'an Jiaotong University, Xi'an, China. He is currently pursuing the Ph.D. degree with Northeastern University, Boston, MA, USA.

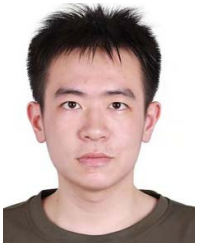

**Yan He** (Student Member, IEEE) received the B.S degree in electronic science and technology from Zhejiang University, Hangzhou, China, in 2018. He is currently pursuing the Ph.D. degree in electrical and computer engineering with Rice University, Houston, TX, USA.

His current research interests include analog and mixed-signal integrated circuits design for power management and hardware security.

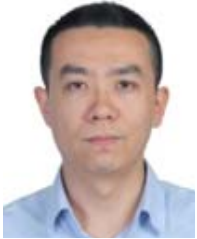

**Jingyu Wang** (Member, IEEE) received the B.S. degree in electronic science and technology, the M.S. and Ph.D. degrees in microelectronics from Xidian University, Xi'an, China, in 2010, 2013, and 2017, respectively.

His current interests include mixed-signal integrated circuits, ADC, image sensors and their applications, biomedical circuits and systems, and RF integrated circuits.

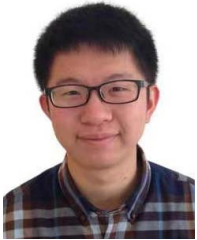

**Sheng Lin** (Student Member, IEEE) received the B.S. degree from Zhejiang University, Hangzhou, China, in 2013, the M.S. degree from Syracuse University, Syracuse, NY, USA, in 2015, and the Ph.D. degree in computer engineering from Northeastern University, Boston, MA, USA, in 2020, under the supervision of Prof. Yanzhi Wang.

His current research interests include privacy-preserving machine learning, energy-efficient artificial intelligence systems, model compression, and mobile acceleration of deep learning applications.

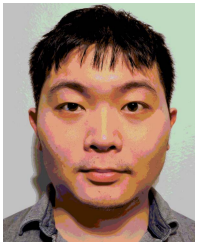

**Dai Li** (Student Member, IEEE) received the B.S. and M.S. degrees in electronics engineering from Tsinghua University, Beijing, China, and the M.S. degree of electrical and computer engineering from Rice University, Houston, TX, USA, in 2010, 2013, and 2017, respectively, where he is currently pursuing the Ph.D. degree.

His research interests include VLSI circuits, hardware security, mixed-signal integrated circuits, and low-power circuits.

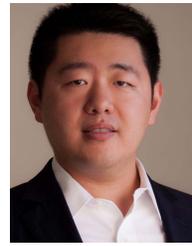

**Yanzhi Wang** (Senior Member, IEEE) received the B.S. degree from Tsinghua University, Beijing, China, in 2009 and the Ph.D. degree from the University of Southern California, Los Angeles, CA, USA, in 2014.

He is currently an Assistant Professor at the Department of ECE, Northeastern University, Boston, MA, USA. His research interests focus on model compression and platform-specific acceleration of deep learning applications. His research maintains the highest model compression rates on representative Deep Neural Networks (DNNs) since 09/2018. His work on Adiabatic Quantum-Flux-Parametron (AQFP) superconducting-based DNN acceleration is by far the highest energy efficiency among all hardware devices. His recent research achievement, CoCoPIE, can achieve real-time performance on almost all deep learning applications using off-the-shelf mobile devices, outperforming competing frameworks by up to 180X acceleration. His work has been published broadly in top conference and journal venues and has been cited above 8500 times.

Dr. Wang has received five Best Paper and Top Paper Awards, has another ten Best Paper Nominations and four Popular Paper Awards. He has received the U.S. Army Young Investigator Program Award (YIP), Massachusetts Acorn Innovation Award, Ming Hsieh Scholar Award, and other research awards from Google, MathWorks, etc. Three of his former Ph.D./postdoc students become tenure track faculty member at the University of Connecticut, Storrs, CT, USA, Clemson University, Clemson, SC, USA, and Texas A&M University-Corpse Christi, Corpse Christi, TX, USA.

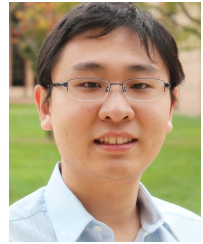

**Kaiyuan Yang** (Member, IEEE) received the B.S. degree in electronic engineering from Tsinghua University, Beijing, China, in 2012 and the Ph.D. degree in electrical engineering from the University of Michigan, Ann Arbor, MI, USA, in 2017.

He is an Assistant Professor of electrical and computer engineering at Rice University, Houston, TX, USA. His research interests include digital and mixed-signal circuits for secure and low-power systems, hardware security, and circuit/system design with emerging devices.

Dr. Yang received the Distinguished Paper Award at the 2016 IEEE International Symposium on Security and Privacy (Oakland), the Best Student Paper Award (first place) at the 2015 IEEE International Symposium on Circuits and Systems (ISCAS), the Best Student Paper Award Finalist at the 2019 IEEE Custom Integrated Circuits Conference (CICC), and the 2016 Pwnie Most Innovative Research Award Finalist. His Ph.D. research was recognized with the 2016–2017 IEEE Solid-State Circuits Society (SSCS) Predoctoral Achievement Award.